\documentclass[ preprint2]{aastex63}

\newcommand{\lya}{Lyman-$\alpha$}

\newcommand{\ergscm}{$\rm erg\;s^{-1}cm^{-2}$}

\newcommand{\ergs}{$\rm\;erg\;s^{-1}$}
\newcommand{\muv}{$\rm\;M_{UV}$}   

\newcommand{\grp}{EGS77}

\newcommand{\obja}{$\rm z8\_SM$}   
\newcommand{\objb}{$\rm z8\_4$}   
\newcommand{\objc}{$\rm z8\_5$}   
\newcommand{\chsq}{$\rm \chi^2$}   

\newcommand{\ed}[1]{{{#1}}}

\shorttitle{Evidence of An Ionized Bubble 680 Myrs after the Big Bang}
\shortauthors{Tilvi  et al.}

\begin{document}

\title{Onset of Cosmic Reionization: Evidence of An Ionized Bubble Merely 680 Myrs after the Big Bang
}

\correspondingauthor{V. Tilvi}
\email{tilvi@asu.edu}
\author{V. Tilvi}
\affiliation{School of Earth and Space Exploration, Arizona State University, Tempe, AZ 85287, USA}

\author{S. Malhotra}
\affiliation{Astrophysics Science Division, Goddard Space Flight Center,  Greenbelt, MD 20771, USA}

\author{J. E. Rhoads}
\affiliation{Astrophysics Science Division, Goddard Space Flight Center,  Greenbelt, MD 20771, USA}

\author{A. Coughlin}
\affiliation{Chandler-Gilbert Community College,  Chandler, AZ 85225-2499 USA}

\author{Z. Zheng}
\affiliation{CAS Key Laboratory for Research in Galaxies and Cosmology, Shanghai Astronomical 
 Observatory, Shanghai 200030, People's Republic of China}

\author{S. L. Finkelstein}
\affiliation{Department of Astronomy, The University of Texas at Austin, Austin, TX 78712, USA}

\author{S.  Veilleux}
\affiliation{Department of Astronomy and Joint Space-Science Institute, University of Maryland, College Park, MD 20742, USA}

\author{B. Mobasher}
\affiliation{Department of Physics and Astronomy, University of California, Riverside, CA 92521 USA}

\author{J. Wang}
\affiliation{CAS Key Laboratory for Research in Galaxies and Cosmology, Department of Astronomy, 
 University of Science and Technology of China, Hefei, Anhui 230026, People's Republic of China}

\author{R. Probst}
\affiliation{NOAO, 950 N. Cherry Avenue, Tucson, AZ 85719, USA}

\author{R. Swaters}
\affiliation{Department of Astronomy and Joint Space-Science Institute, University of Maryland, College Park, MD 20742, USA}

\author{P. Hibon}
\affiliation{European Southern Observatory, Alonso de Cordova 3107, Casilla 19001, Santiago, Chile}

\author{B. Joshi}
\affiliation{School of Earth and Space Exploration, Arizona State University, Tempe, AZ 85287, USA}

\author{J. Zabl}
\affiliation{Univ Lyon, Univ Lyon1, Ens de Lyon, CNRS, Centre de Recherche Astrophysique 
 de Lyon UMR5574, F-69230 Saint-Genis-Laval, France}

\author{T. Jiang}
\affiliation{School of Earth and Space Exploration, Arizona State University, Tempe, AZ 85287, USA}

\author{J. Pharo}
\affiliation{School of Earth and Space Exploration, Arizona State University, Tempe, AZ 85287, USA}

\author{H. Yang}
\affiliation{School of Earth and Space Exploration, Arizona State University, Tempe, AZ 85287, USA}

%

%

%
%
%
%
%


\begin{abstract}

While most of the inter-galactic medium (IGM) today is permeated by ionized hydrogen,
 it was largely filled with neutral hydrogen for the first 700 million years after the Big Bang. 
  The process that ionized the IGM (cosmic reionization) 
is  expected to be  spatially inhomogeneous,
   with fainter galaxies playing  a significant role.
    However, 
   we still have only a few direct constraints on the reionization process.  Here we report the first 
   spectroscopic confirmation of  \ed{ two galaxies  and very likely  a third galaxy} 
   in a group (hereafter \grp) 
   at redshift $z = 7.7$, merely 680 
   Myrs after the 
   Big Bang. The physical separation among the three members  is $<$ 0.7 Mpc.
   We estimate the radius of ionized bubble of the brightest galaxy to be about 1.02 Mpc, and 
    show that the individual ionized bubbles formed by all three galaxies likely 
   overlap significantly, forming a large yet localized ionized region, which leads to the spatial 
   inhomogeneity in the reionization process. 
   \ed{It is striking that two of three galaxies in \grp\ are quite faint
in the continuum, 
     thanks to our selection of reionizing 
    sources using their \lya\   line emission. 
Indeed, one is 
    the faintest spectroscopically confirmed galaxy yet discovered
     at such high redshifts. }
    Our observations provide direct constraints 
    in the process of 
    cosmic reionization, and allow us to  investigate the properties of sources responsible for
     reionizing the universe.

\end{abstract}



\section{Introduction} \label{sec:intro}

Cosmological   simulations indicate that the process of reionization, which is 
expected to be patchy or spatially inhomogeneous  (Furlanetto et al 2004, 
Iliev et al. 2006, Zahn et al. 2007, Mesinger A. et al. 2008, Jensen, H. et al 2014),
  started when 
%
 intense UV radiation from  individual galaxies or groups of galaxies 
first  ionized their local surroundings, and formed local 
 ionized bubbles. 
 These ionized bubbles later grew to fill the 
 entire IGM, marking the end of the reionization process.

Star-forming galaxies at high redshifts are expected to have 
contributed to the 
reionization process \citep[e.g.][]{sti04,bou15}.
These same galaxies, via their \lya\ emission,   
provide a practical tool to study  the reionization process$-$ the evolution of neutral hydrogen in 
the IGM.
Currently however, we lack  a clear observational evidence of the 
spatial inhomogeneity or  the  sources responsible for 
the cosmic reionization.
Therefore,  to  study the nature of cosmic reionization, and quantify 
 galaxies' contribution to this  process, 
we have carried out a unique narrow-band (NB)  imaging survey,  
the  Cosmic Deep And Wide Narrowband   survey
(DAWN; PI: Rhoads)
to observe \lya\ emitting galaxies
at high redshifts. 
The NB technique has been proven successful in identifying $z>6$ galaxies \citep{hu10,ouc10,rho12}
which
otherwise could  go  undetected in the traditional  broad-band selection techniques \citep[e.g.][]{ste03}. 
Here we present discovery of the most distant galaxy group, identified using NB imaging, 
and  confirmed via spectroscopic observations. 

  \begin{figure*} [t]
  \centering
\epsscale{1.15}
\plotone{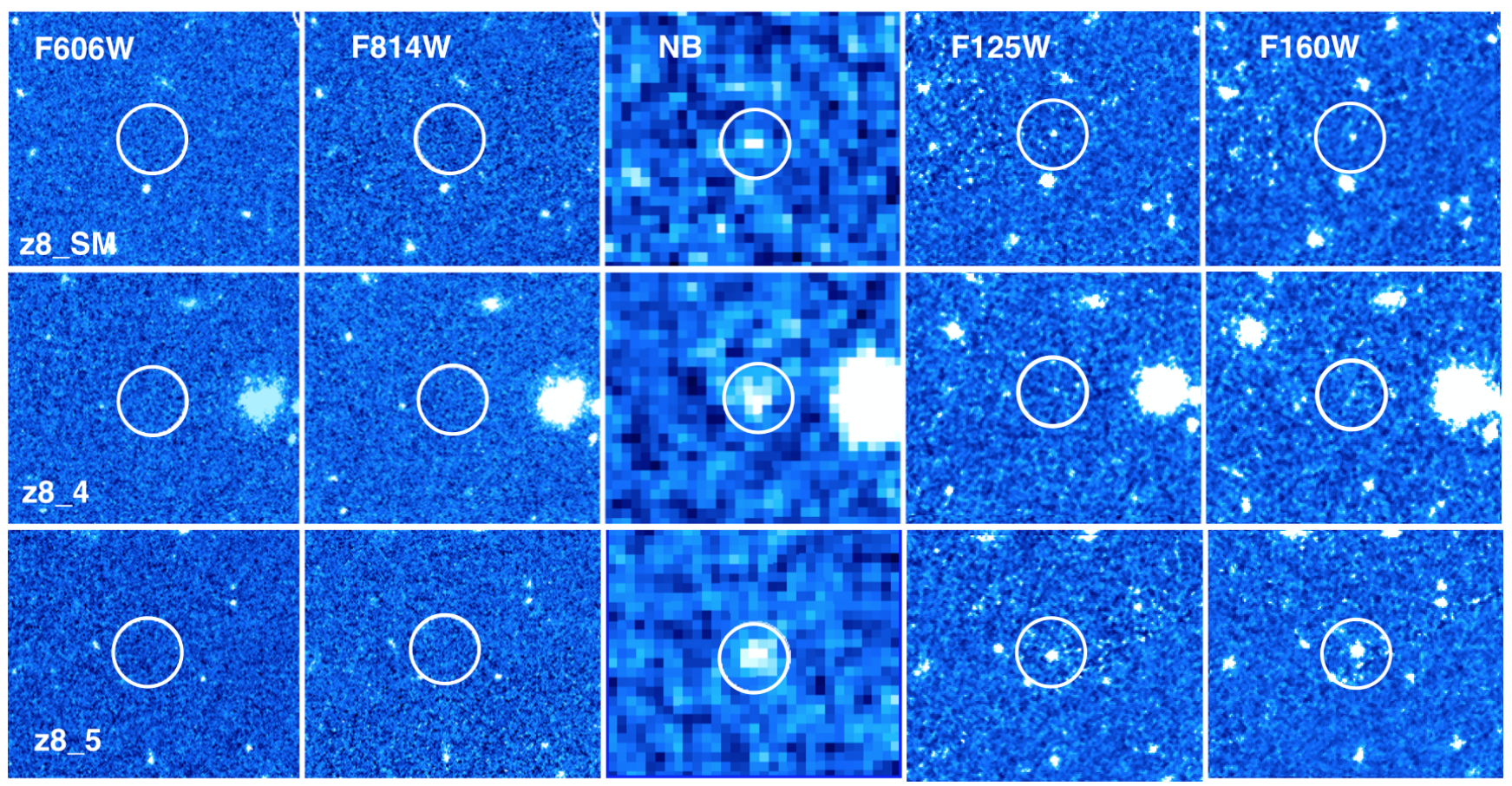}
\small{
\caption{
\textbf{Image  cutouts of all three members of \grp.} 
All galaxies are  
detected at redder wavelengths (NB, F125W, and F160W), while undetected at the visible 
wavelengths  (F606W and F814W).
The sharp drop in the flux at visible wavelengths is consistent with the  objects being at  redshifts $z\geq7$.
All cutouts are $\sim$ 12\arcsec\ on the side, and the circles enclosing objects have 1.4\arcsec\ radius.
NB images appear pixelated due to the coarser pixel resolution (0.4\arcsec\ per pixel) compared with the much finer
 Hubble Space Telescope (HST)  resolution (0.06\arcsec\ per pixel). 
Bright fluxes in the NB images indicate the presence of strong emission lines.
\vspace{0.3cm}
}
}
\end{figure*}

\section{Imaging}
\subsection{Observations \& Data Reduction}
We obtained deep NB imaging observations of the 
Extended Groth Strip (EGS) field (RA 14:19:16 DEC +52:52:13), 
 as part of the DAWN survey.
 This is a uniquely deep survey given its sensitivity as well as 
 area coverage, with a primary 
objective of  identifying  galaxies
 at  redshift $z=7.7$.
 Here we present relevant details of the  DAWN survey (Rhoads et al in prep, Coughlin et al 2018).
The DAWN    survey was carried out using a
  custom built  narrow-band filter (FWHM=35\AA,  central wavelength= 10660\AA), mounted
  on the NOAO Extremely Wide-Field InfraRed Imager
   (NEWFIRM)\citep{pro04, pro08}
   at the Kitt Peak 4m Mayall telescope.
The NEWFIRM instrument  has a wide field of view (28x28 arcmin$^2$) with a resolution of 0.4\arcsec\ per pixel.
We obtained individual images  with  600s integration time and  Fowler samples of 16
with   8 digital averages  during the readout. 
To achieve clean sky background
subtraction we used random dithering  with a  dither size of 45\arcsec.
 
 The data reduction was  primarily done  using the NEWFIRM science data reduction pipeline \citep{swa09}.
 However, for generating the final stack of all the images produced by the pipeline (sky subtracted, cosmic-rays
 cleaned, re-projected) we used our own scripts to remove bad frames that were visually 
 inspected in order to maximize the   signal-to-noise  ratio of astronomical objects.
 The final NB stack is equivalent to a total integration time of 67 hrs,  yielding    $5\sigma$  line flux sensitivity
 of $\sim 7 \times10^{-18}$\ergscm.
In addition to the NB image described above, to select high-redshift \lya\ emission line candidates, 
we also used
publicly  available broadband images at visible  wavelengths  (HST/ACS F606W, F814W)  and at 
near-IR wavelengths (HST/WFC3 F125W, F160W).
The observations at visible wavelengths were taken as part of the GO10134 program (PI: M. Davis).
Near-IR and IRAC observations were taken as part of the GO12063 program (PI: S. Faber), and 
GO61042 program (PI: G. Fazio), respectively.

\subsection{Selection of $z=7.7$ Candidate Galaxies}
To generate the source catalog we first aligned all images that include   
HST/ACS images (F606W, F814W), 
near-IR  NB1066 image (NB from DAWN survey), 
and near-IR images (F125W, F160W)
 onto
a common world coordinate system grid. We then used a source detection software  
(SExtractor; Bertin \& Arnouts 1996)
in dual image mode where the 
detection image (in this case the NB) is used to identify the pixels
associated with each object, while the fluxes are measured from
the respective  photometry image. 

For robust selection of candidate galaxies 
at $z=7.7$ we followed a set of criteria 
that has  yielded a high spectroscopic success
rate at $z$ = 4.5 and 5.7\citep{rho01, rho03, daw04, daw07, wan09}. 
Following this,  each of our  candidates had to satisfy
all of the following criteria:
1) 5$\sigma$ detection in the  NB filter, 
2) 3$\sigma$ significant narrowband excess (compared to the F125W
image),  and 
3) non-detection ($< 2\sigma$)  in the individual optical images (F606W, F814W). 
Criteria 1 \& 2  ensure real emission line sources while  criterion 3
eliminates most low-redshift sources.
Using this set of criteria we identified three $z=7.7$ candidates:  \objc, \objb, and
\obja\   for spectroscopic
follow-up. 
Formally, \obja\ has a NB  S/N$\sim$ 4 (just below our selection
 threshold), however  it has an aggregate S/N=13 in NB+F125W+F160W. 
 Given that it
 satisfied all other selection criteria, and given its
 proximity with other two galaxies,  we included it for the spectroscopic follow-up observations.
It should be noted that galaxy \objc\ was independently identified previously by Oesch et al (2015) using 
Lyman-break selection 
 technique.
%
%

Figure  1 shows image cutouts of all three candidates in five filters.
All three  galaxies  have significant fluxes 
 in the NB filter, indicating the  presence of strong 
emission lines, most likely \lya\ lines  at the observed wavelength of $1.066\mu m$.
None of the galaxies have detectable fluxes 
at the visible wavelengths suggesting that these galaxies are consistent with
being at redshifts $z\geq7$.
All three galaxies are detected in both F125W and F160W (see Table 1), \ed{despite
two of them being faint,  }
%
making \objb\  the faintest galaxy discovered at such high redshifts,
thanks to the NB selection technique in which detection of galaxies 
 does not depend on the continuum brightness.

\subsection{Photometric Redshifts And Spectral Energy Distributions}
To measure the photometric redshifts, we made use of 
spectral energy distribution (SEDs) templates.
We obtained the best-fit  SEDs using   $EAZY$\citep{bra08}
which provides photometric redshift probability distribution $p(z)$
by finding the best-fitting combination of redshifted galaxy spectral templates 
to the observed photometry.
Because our galaxy sample is selected to have emission lines, to derive the best-fit SEDs and
photometric redshifts,  we chose a set of spectral templates that have emission lines (eazy\_v1.3).
These templates are corrected for the intergalactic absorption by neutral hydrogen, following  the 
Madau prescription\citep{mad95}.
We allowed a wide range of redshift grids ($z=0.1$ to $   z=9$) to search for the best-fit SED template.
\ed{In addition to the photometry (discussed in Section 2.2),   we also used HST/WFC3 F105W 
photometry (GO:13792, PI: Bouwens) for \objc\ and 
CFHT-Y band photometry for \objb\ and \obja.}

As shown in Figure 2, for  all three galaxies  the best-fit  SEDs (shown in blue color)
 prefer  spectral templates that correspond
 to a photometric redshift $z_{phot}=7.7$.
 This is  also evident from the   $p(z)$ 
 (shown in the  inset); the presence of strong emission lines in the NB filter yield  very tight constraints
 on the $p(z)$.
For completeness, we also show the  low-redshift galaxy templates (grey color), which however
 are disfavored due to much larger \chsq\  values of the SED fit.
For the brightest galaxy \objc\ (previously identified in Oesch et al 2015), 
the difference  between  best-fit \chsq\ value of low-redshift template and 
high-redshift  template is 86. For the remaining galaxies \objb, and \obja\, while 
the difference between  best-fit \chsq\ value of low-redshift template and 
high-redshift  template is $> 1$, it is not as high as for \objc\  given their lower S/N.
Furthermore, all three galaxies have large  rest-frame \lya\ equivalent widths  $\rm EW_{rest}>23$\AA\ 
(see  Table 1), which likely makes them visible as \lya\ emitting galaxies.

  \begin{figure*} [b!]
  \centering
\epsscale{1.15}
\plotone{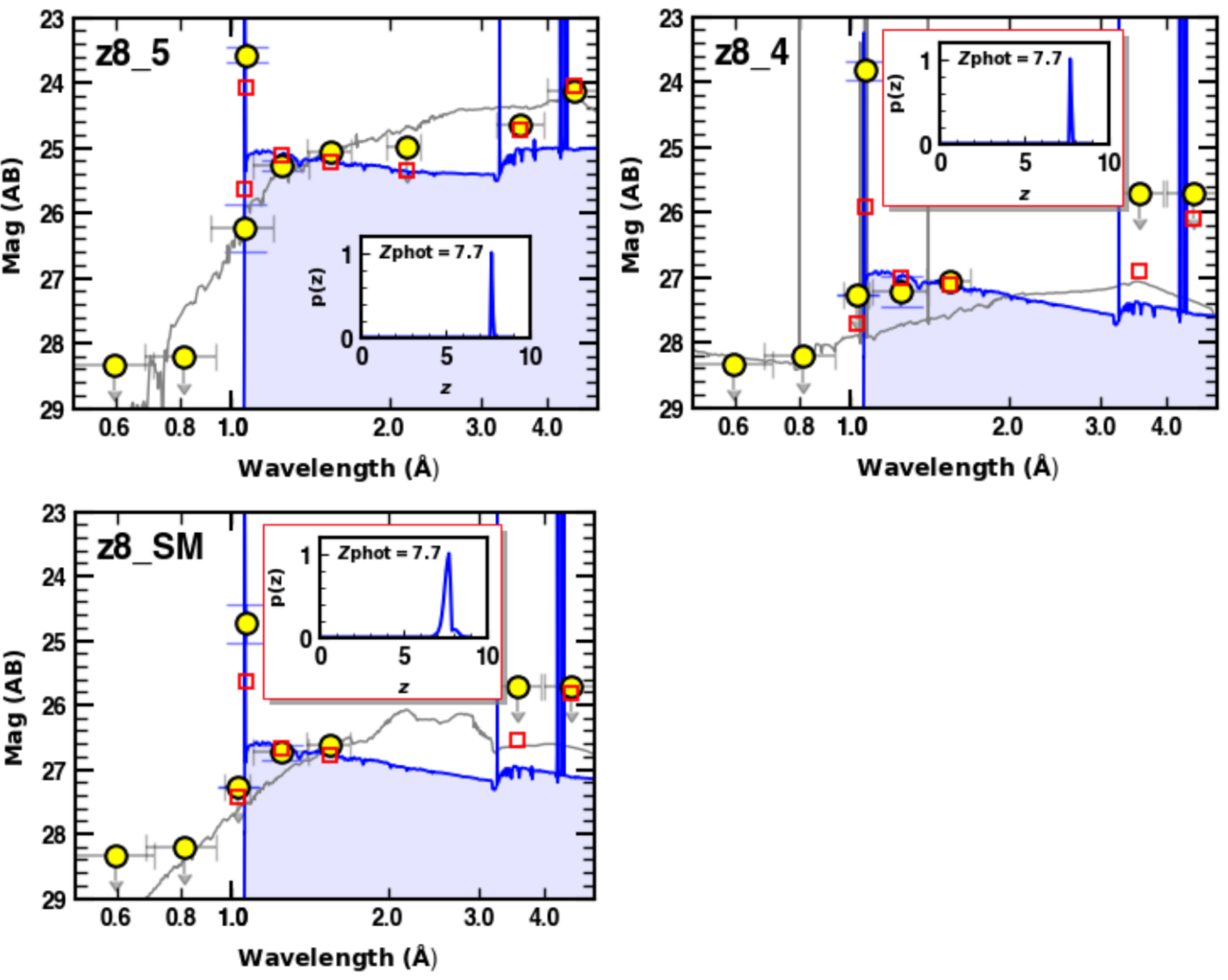}
\vspace{-0.2cm}
\small{
\caption{
\textbf{Spectral energy distributions (SEDs) of \grp:} SEDs  based on fits to 
the photometry obtained from HST/ACS (F606W, F814W), NEWFIRM (NB), 
HST/WFC3 (F125W, F160W), and Spitzer IRAC (ch1, ch2).
\ed{In addition, \objc\ photometry includes HST/WFC3 F105W observations 
while \obja\ and \objb\ include
CFHT-Y band photometry.}
The blue line  represents the best-fit spectral template while the filled yellow circles
represent photometric observations. Open squares indicate best-fit template
fluxes convolved with the respective filters.
Circles with downward pointing arrows represent $2\sigma$ non-detection limits.
The best-fit photometric redshift distribution  (shown in the inset) yields  $zphot=7.7$ which
implies that there is a high probability of these galaxies being at high redshifts.
For completeness, we also show  low redshift galaxy spectral templates (shown in grey color) 
which are disfavored
given their larger  $\chi^2$ values.
}
}
\end{figure*}

\begin{deluxetable*}{cccccccc}
\tablenum{1}
\tablecaption{Photometry  and spectroscopic measurements of \grp}
\tablewidth{0pt}
\tablehead{
\colhead{ID} 	& \colhead{RA}		& \colhead{DEC} 	& \colhead{F606W} 	& \colhead{F814W} & \colhead{NB} & \colhead{F125W} & \colhead{F160W}\\
\colhead{} 	& \colhead{J2000}	& \colhead{J2000}	& \colhead{mag	}	&\colhead{mag}	 & \colhead{mag}   & \colhead{mag}	& \colhead{mag}
%
%
}
\startdata
 z8\_{SM}   	&  14:20:35.694      	& +53:00:09.318	& $<$ 28.3$^1$ & $<$ 28.2$^1$  & 24.76$\pm$0.35	& 26.76$\pm$0.13	& 26.66$\pm$0.11  	 \\
z8\_4   		&  14:20:35.169     	& +52:59:40.613	& $<$ 28.3$^1$  & $<$ 28.2$^1$ & 23.85$\pm$0.15	& 27.25$\pm$0.21	& 27.10$\pm$0.16 	  \\
z8\_5   		&  14:20:34.872      	& +53:00:15.242	& $<$ 28.3$^1$ &$<$  28.2$^1$  & 23.60$\pm$0.12	& 25.29$\pm$0.03	& 25.08$\pm$0.03	   \\
\hline 
\\
\multicolumn{8}{c}{Spectroscopic measurements } \\
\hline
%
\colhead{ID} 	& \colhead{$z_{spec}$}   &\colhead{$f_{Ly\alpha}$ }&  \colhead{EW$_{rest}$} & \colhead{L$_{Ly\alpha}$}  & \colhead{HII radii $^2$}	& \colhead{SNR}   & \colhead{Distance$^3$} \\
\colhead{}  & \colhead{}   &\colhead{{(\ergscm)} } &  \colhead{\AA} & \colhead{$10^{43}$(\ergs)}  & \colhead{pMpc} &  \colhead{$\rm (Ly\alpha)$} & {Trans$\;$}  {LoS} \\
\hline
   z8\_{SM}  	&  7.767      &{0.29$\pm$0.06} $\times10^{-17}$	& 23$\pm$6 		& 0.2$\pm$0.1		&{ 0.55}	  & 4.9	& 0.06  $\;$ 0.9\\
   z8\_4   		&  7.748    & {0.56$\pm$ 0.09} $\times10^{-17}$	   & 71$\pm$18	& 0.4	$\pm$0.1		& {0.69}	  & 6.0 	&0.09 $\;$ 0.2\\  
   z8\_5   	&  7.728      	&{1.7$\pm$ 0.14} $\times10^{-17}$		  & 37$\pm$3		& 1.2	$\pm$0.1		&  {1.02}	  & 12.1	& 0.08 $\;$ -0.5\\
\enddata
\tablecomments{
$^1$$2\sigma$ magnitude  limits. All magnitudes are AB magnitudes.\\
$^2$HII bubble radii based on Fig 15 from Yajima et al 2018.\\
$^3$Trans and LoS are the transverse and line-of-sight separation of each galaxy from the 
flux-weighted mean location of the group center, measured in pMpc. 
}
\end{deluxetable*}


\section {Spectroscopic Observations}
To confirm the  photometric redshifts of these galaxies, 
 we performed Y-band spectroscopy using 
 Multi-Object Spectrometer for Infra-Red Exploration (MOSFIRE)
 spectrograph (McLean et al. 2012)
%
on the Keck I telescope.
 The MOSFIRE instrument allows us to obtain spectroscopic
observations of multiple objects simultaneously, 
with the Y-band covering \lya\  lines redshifted to \ed{$z=7-8.2$.}
 We targeted three  \lya\ galaxy candidates in the EGS field, 
 as our primary science targets, and used low-redshift emission line candidates
 as fillers.
Observations were taken on May 06, 2017, with each individual exposures
having  140 sec integration time and  $AB$   pattern dither offsets
along the slit with offset of $\pm$1\arcsec\  from the center. 
 The spectroscopic conditions were good with a typical seeing of $\sim 0.7$\arcsec, and 
a total integration time of about 4 hrs per object.

We  reduced data using  the public 
MOSFIRE data reduction pipeline. It performs standard data reduction 
procedures including  sky subtraction, rectification of the 2D spectra, and
wavelength calibration. 
For a given object,  it  also produces corresponding
signal-to-noise ratio (SNR)  and a sigma  image.
The final 2D science spectra has a spatial resolution of 0.18\arcsec\ per pixel and
a dispersion of 1.086 \AA\ per pixel.
For absolute flux calibration of 1D spectra, we compared the measured line flux of
\objc\ with the calibrated flux from \citet{oes15},
and converted instrumental
flux (in counts)  to the absolute flux.  The conversion factor derived for \objc\ is  then 
used to calibrate spectra of  \obja\  and \objb\ .

  \begin{figure*}[t]
  \centering
\epsscale{1.15}
\plotone{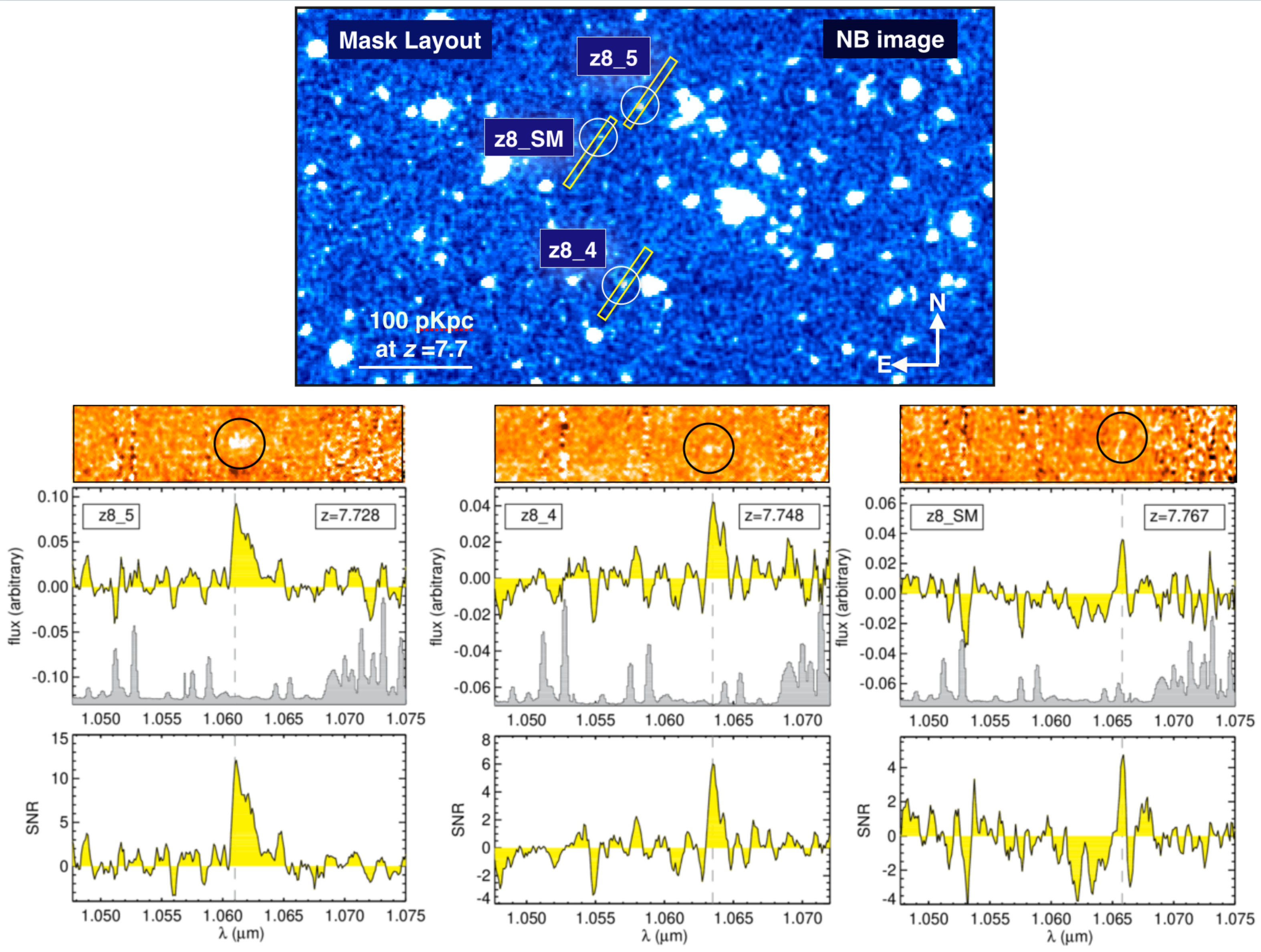}
\small{
\caption{
\textbf{Spectroscopic observations of \grp.}
Top: 
A portion of the Extended Groth Strip (EGS)  field centered on \grp,
and the multi-object mask layout used  for MOSFIRE Y-band spectroscopic observations.
 The mask is overlaid on the NB image \ed{ ($\sim$ 2.3\arcmin $\times$1.2\arcmin) }where the image is slightly smoothed for  clarity.
Lower panels: Top and middle rows show final stacked 2D and 1D spectra respectively, 
 while bottom row
shows signal-to-noise ratio (SNR). 
\ed{The 
 maximum of the SNR is normalized to the total SNR of the \lya\   line.}
Shaded grey region shows night sky lines with arbitrary normalization. 
\lya\ lines (shown by black circles,  and represented with vertical dashed lines in the middle row) 
are  detected in all three galaxies
at the expected observed wavelength of 
about $1.066\mu m$. 
}
}
\vspace{0.3cm}
\end{figure*}
%

Figure 3 shows the  final co-added two-dimensional (2D) and one-dimensional (1D) 
Y-band spectra of all three galaxies. 
\ed{As seen, both galaxies \objc\ and \objb\ show  prominent emission lines(S/N $>$5) while
\obja\ has S/N=4.9. } In the following we show that these 
  are very likely
\lya\ emission at $z=7.7$.


For both galaxies \objc\ and \objb, emission  lines are 
free from  the night sky OH lines, as shown in 
shaded region (middle row).
For  \obja, while its  emission line is close to  a faint  night sky line, 
given its higher S/N compared to the  night sky lines,  
 it is very likely  a genuine  emission line.  
Furthermore, the observed Y-position of the  emission line in the 2D 
spectrum of \obja\  matches well with that expected based on the 
slit position in the mask (Fig 3 top panel), supporting our conclusion that this is 
a real emission line.
%
%
%
%
\ed{We note that  the presence of a faint night sky line at the position of the \lya\   
line may possibly affect the \lya\ line flux measurements. 	
However, given the  faintness of the night sky line, we expect that it will have a minimal impact 
on the measurement of the \lya\  line flux. }
In the following we demonstrate  that  these are very likely  \lya\  lines at $z=7.7$.

One of the characteristics that distinguishes  \lya\ emission from star-forming galaxies
 and other emission lines is the line  asymmetry\citep{rho03}
 or 
 \textit{Skewness}\citep{kas06}.
%
%
To quantify  the  \textit{Skewness} in the line, we calculated the weighted skewness
parameter $Sw$, 
and found $Sw=22\pm6$\AA\  for \objc.
This confirms  that \objc\  is  a $z=7.7$ galaxy because $Sw>3$\AA\ 
is not seen in low-redshift emission lines of [OII], [OIII], and H$\alpha$ \citep{kas06}. 
Indeed   \objc\  galaxy  was  previously 
identified as a Lyman break candidate (labelled as EGS-zs8-1), and spectroscopically 
confirmed  as  a $z=7.7$ galaxy\citep{oes15}.
  Furthermore, recent H-band spectroscopic observations of this galaxy show presence of 
  [CIII] 1909 doublet\citep{sta17}.
Thus, given all the evidence, \objc\ is unequivocally a \lya\ emitting galaxy at $z=7.7$.
 For  galaxy \objb\ the weighted skewness
parameter for the observed emission line is $Sw=17\pm7$\AA\, confirming the line to be 
\lya\  at $z=7.7$.
%
For  \obja, 
 we can not reliably measure the asymmetry of the line 
 given  its    lower S/N in the spectrum.
%
%
%
%
%
However, based on the  best-fit SEDs,   both  \objb\ and \obja\  favor high redshift solutions. 
Furthermore,  if these were faint, low-redshift galaxies, the best-fit low redshift SEDs 
implies a clear detection in F606W and F814W filters.
Thus, given all the evidence,  \ed{  \objc\ and   \objb\ are unequivocally at redshifts   
$z=7.728$ and  $z=7.748$ respectively,
 and \obja\ is also very likely at redshift $z=7.767$.}
%
%
%
%
%
%

  \begin{figure*}[t]
  \centering
\epsscale{1.15}
\plotone{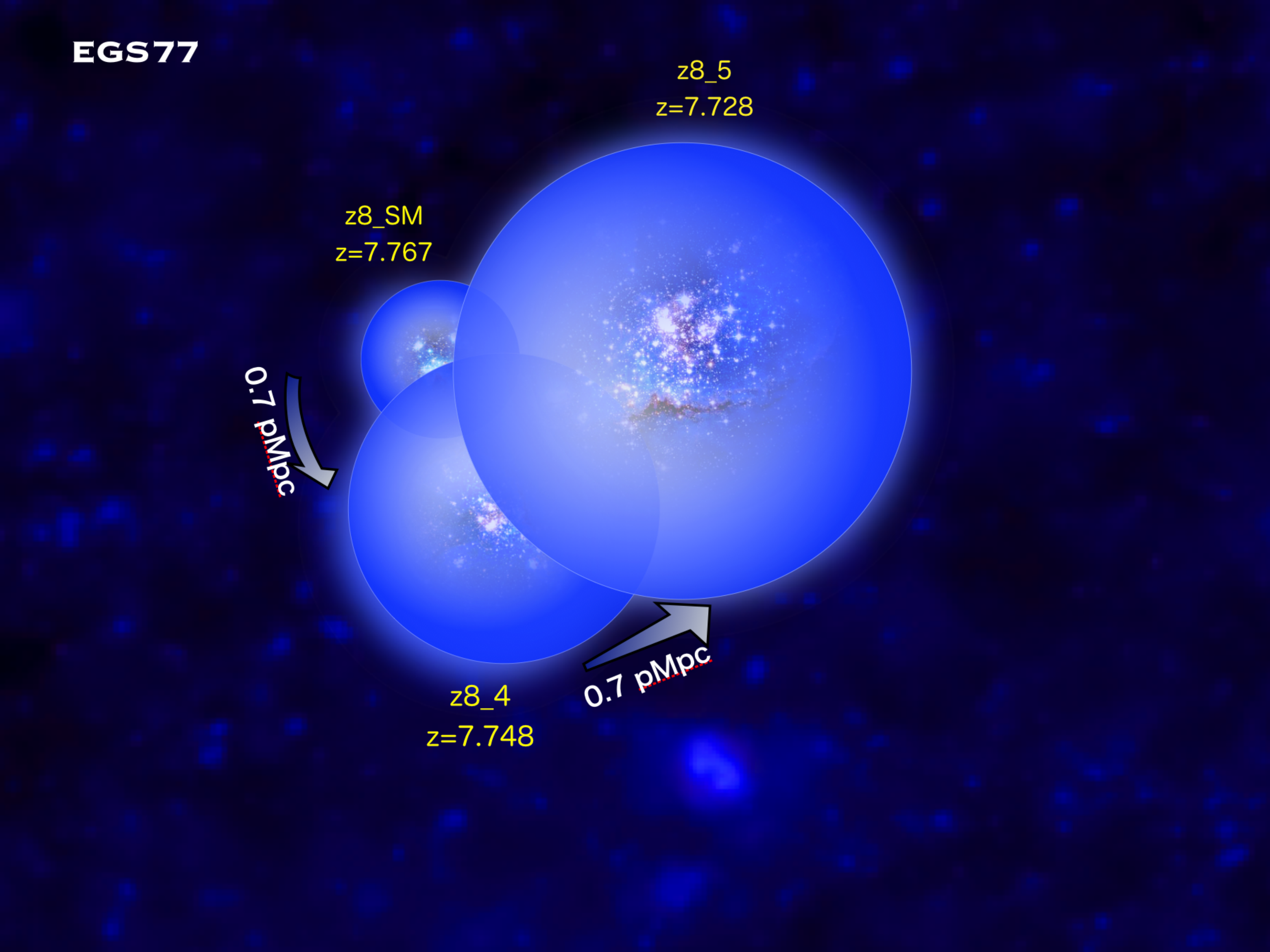}
\caption{
\small{
\textbf{Rendition of the galaxy group EGS77: }
Rendition showing ionized 
bubbles formed by three galaxies near redshift 7.7. Galaxy \objc\ is the brightest 
of all three, with the largest ionized bubble, and lies in the front. It is estimated to
 produce an ionized bubble of radius 1 pMpc, which will help clear the way for
  transmission of \lya\  photons from the two fainter galaxies. The physical 
  separation between each pair of galaxies is merely 0.7 pMpc along the line of 
  sight, which implies that their ionized bubbles overlap substantially to create a 
  continuous path allowing \lya\  photons to escape.
\vspace{0.3cm}
}
}
\end{figure*}

\ed{It is striking that while  \objb\ and \obja\  are faint (\muv $> -20.3$ mag ), 
both  show \lya\  emission lines. Moreover, despite low 
number density of such faint galaxies at $z >7$, all three galaxies are spectroscopically 
confirmed. This high spectroscopic success rate is likely because (1) our NB technique preselects galaxies with detectably strong line emission, and 
(2) \grp\  is likely to have formed a large ionized bubble, allowing \lya\   
photons to escape. We discuss this in more detail in the following section.}

\section{Visibility of Lyman-${\alpha}$}
The visibility of \lya\ emission from star-forming galaxies at high redshifts 
depends on several factors 
including star-formation rate,  ionizing photon budget, galactic outflows, and the density of 
neutral hydrogen surrounding the galaxy. The star-formation rate and the ionizing photon
budget will directly influence the amount of ionized gas forming an ionized bubble which
in turn  allows \lya\ photons to travel unattenuated along the line-of-sight\citep{mal06}.

The separation between the most distant member \obja\ ($z=7.767$) 
 and \objb\ ($z=7.748$) along the line-of-sight  is 
about 0.7 pMpc (physical
 Mpc), which is  same as the  separation between \objb\ and \objc\ ($z=7.728$).
 In the transverse direction (i.e. projection on the sky) all three members are
 much closer to each other.
 The separation 
between  \obja\ and \objc\  in the transverse direction 
 is  10\arcsec ($0.05$ pMpc), while
 the separation between  \objb\ and  \objc\ is 
 35\arcsec (0.18 pMpc).
%
Thus, given the proximity of all three galaxies with each other in both   the transverse 
direction and
along the line-of-sight, 
these galaxies will   form a continuous 
attenuation-free path for \lya\ photons if the radii of their ionized bubbles 
are $\geq$ 0.35 pMpc
along the line-of-sight direction (Fig. 4).
This is because the ionized region is large enough that the 
\lya\ photons are redshifted by the time they reach the neutral hydrogen
boundary, and thus can escape.
This is supported by 
recent spectroscopic observations of the brightest galaxy \objc,
showing a  [CIII] 1909 doublet\citep{sta17}, 
which yields 
a velocity offset $\Delta V_{Ly\alpha} $=340$^ {+15} _{-30} \; \rm km \; s^{-1}$.
When compared to a FWHM=360 $\rm km \; s^{-1}$  for this line\citep{oes15}, 
it implies that a substantial fraction of \lya\ 
photons are leaving the galaxy at 
340- 520  $\rm km \; s^{-1}$.

\subsection{Estimation of Bubble Sizes}
We now estimate the sizes of ionized bubbles formed by these galaxies, based on  theoretical 
model  where
the relation between \lya\ luminosities and bubble sizes 
has been predicted through simulations, while   the star
 formation rates  is derived using growth rate of halo mass with a
  constant tuning parameter \citep{yaj18}. 
%
 The growth rate of halos is calculated using the halo merger trees
based on an extended Press-Schechter formalism \citep{som99,kho01}. 
These simulations 
 reproduce the observed star formation
rate density as well as the UV luminosity function of Lyman break selected galaxies
at $z\sim7$ and $z\sim8$ \citep{bou12}. 
Next, to  estimate the   sizes of ionized bubble which is proportional to the ionizing photon budget, they
use star formation history of each halo using stellar population synthesis code  STARBURST99 \citep{lei99}. 
Finally, the \lya\ luminosity of each galaxy is calculated based on the number
of ionizing  photons  
absorbed within the galaxy.  Photons that are absorbed  will produce \lya\ 
photons while photons that escape cause the cosmic reionization.
%
%
%
%
%
%
%
%
%
Based on these simulations (Figure  15 from Yajima et al 2018),  
the luminosity of our brightest galaxy 
 \objc\  yields a  radius of 
 1.02 pMpc for the HII bubble. 
 For the remaining two galaxies  \objb\ and  \obja\ we get
the bubble sizes of  0.69 and 0.55 pMpc respectively (see Fig 4).
%
This implies that the ionized bubbles produced by individual  galaxies 
overlap significantly, forming a continuous non-attenuating path through which \lya\ photons can easily
escape and hence  be observed\citep{lar18}. 
This is currently the { \it only}  observational evidence of an ionized bubble 
formed by a group of three galaxies (\grp)  during the onset of cosmic reionization.
This localized ionized region leads to the expected spatial  inhomogeneity in the reionzation
process, and provides evidence of fainter galaxies' contribution to this process.

\section {summary}
In this \textit{Letter} we reported the discovery of the farthest galaxy group \grp\ at $z=7.7$, 
merely 680 Myrs after the Big Bang. 
\grp\ was  initially
identified using NB imaging observations from the DAWN survey, and later confirmed 
via spectroscopic observations using MOSFIRE spectrograph on the Keck telescope.
\ed{It is striking that all three galaxies in \grp\ are spectroscopically confirmed via \lya\ 
emission line} ($\rm SNR_{Ly\alpha} \gtrsim 5$) \ed{despite two of them being faint in the continuum.
In fact \grp\ contains the faintest galaxy (in terms of the continuum brightness) discovered and spectroscopically confirmed at this 
redshift.} Based on model  and simulations from
 literature, we found that the ionized bubbles produced by all three galaxies overlap
 significantly, producing a large, yet localized ionized region giving rise to the 
 expected inhomogeneity in the reionization process.
 Thus, this is the first observations of a galaxy group responsible for the cosmic reionization.
 The  \textit{James Webb Space Telescope} will be sensitive to the rest-frame UV continuum of
such fainter galaxies in  the reionization epoch, and therefore can provide a much larger
sample of galaxies responsible for  the re-ionization process. 
Furthermore, 
future 21cm observations from SKA-2\citep{dew09} will be able to  probe the  ionized structures 
with angular size as large as  $\sim10$\arcmin,
similar to the size of ionized bubble produced by \objc\ at $z=7.7$.

\acknowledgements  We thank the US National Science Foundation 
 (NSF  Grant AST-1518057),  and NASA for its financial support through the 
 WFIRST Science Investigation Team program (NNG16PJ33C).
 Z.Y.Z. thanks NSF of 
 China (11773051) and the CAS Pioneer Hundred Talents Program.
 Data presented herein were obtained at the W.M. Keck Observatory, which is 
 operated as a scientific partnership among  Caltech, the 
 University of California and the NASA. 
 The Observatory was made possible by the generous financial support of the W.M. 
 Keck Foundation.  The authors wish to recognize and acknowledge the very 
  significant cultural role and reverence that the summit of Maunakea has always had within 
  the indigenous Hawaiian community. We are most fortunate to have the opportunity to conduct 
  observations from this mountain.






{}



\end{document}